\documentclass[conference]{IEEEtran}
\IEEEoverridecommandlockouts
\usepackage{cite}
\usepackage{amsmath,amssymb,amsfonts}
\usepackage{algorithmic}
\usepackage{graphicx}
\usepackage{textcomp}
\usepackage{xcolor}
\usepackage{tabularx}
\def\BibTeX{{\rm B\kern-.05em{\sc i\kern-.025em b}\kern-.08em
    T\kern-.1667em\lower.7ex\hbox{E}\kern-.125emX}}
\begin{document}

\title{Emotion Recognition from Physiological Signals Using Machine Learning Algorithms Under Controlled Emotional Stimuli}

\author{\IEEEauthorblockN{1\textsuperscript{st} Aditi Site}
\IEEEauthorblockA{\textit{Data Science Research Centre} \\
\textit{Tampere University}\\
Tampere,Finland \\
aditi.site@tuni.fi}
\and
\IEEEauthorblockN{2\textsuperscript{nd} Annariina Lohiranta}
\IEEEauthorblockA{\textit{Data Science Research Centre} \\
\textit{Tampere University}\\
Tampere,Finland \\
annariina.lohiranta@tuni.fi}
\and
\IEEEauthorblockN{3\textsuperscript{rd} Tarmo Lipping}
\IEEEauthorblockA{\textit{Data Science Research Centre} \\
\textit{Tampere University}\\
Tampere,Finland \\
tarmo.lipping@tuni.fi}
}

\maketitle

\begin{abstract}
Emotion recognition using physiological signals plays a crucial role in well-being analysis, affective computing and human–computer interaction. This study investigates the performance of multiple machine learning models in classifying targets such as discrete emotions with varying granularity, valence and arousal using physiological signals such as Electrocardiogram (ECG) and Galvanic Skin Response (GSR). In here, we extracted various time- and frequency-domain features from the ECG and GSR data to train machine learning models. The results indicate that categorizing discrete emotions with fewer emotions and categorizing arousal achieves good classification accuracy with tree-based models. XGBoost achieved accuracy of 52.8\% for classifying discrete emotions and Random Forest achieved accuracy of 53.6\%  for classifying arousal. In both the cases, the combination of ECG+GSR feature sets achieved best performance. These findings highlight the effectiveness of physiological signals in capturing emotional states and support their use for emotion recognition systems.
\end{abstract}


\section{Introduction}
Emotional health is directly related to mental and physical health. To maintain a balance between mental/physical health and overall well-being, it is essential to be able to understand and regulate emotions \cite{Intro_1}. Automatic emotion recognition has become an important research area with applications in human–computer interaction, healthcare, and analyzing overall well-being. Emotion recognition can be performed either using behavioral cues or physiological data. Behavioral cues including facial expressions, speech and words, can be consciously manipulated and are sensitive to cultural differences. In contrast, physiological signals such as heart rate, brain activity skin temperature and skin conductivity provide objective measurements of autonomic nervous system activity and are less susceptible to voluntary control making them more reliable for emotion recognition task.

Recent studies focus on the use of physiological signals for emotion recognition tasks \cite{Intro_2}. Commonly used physiological signals are electrocardiogram (ECG), galvanic skin response (GSR), electroencephalogram (EEG), photoplethysmogram (PPG) and skin temperature (ST) \cite{Intro_3}. ECG signals measure electrical activity of the heart and features such as heart rate, heart rate variability are particularly effective in capturing arousal-related emotional responses. GSR signals provide direct measures of emotional arousal through sympathetic activation especially for the intense emotions such as sad, anger, fear etc. EEG signal captures electrical activity of the brain and measures emotional processing and cognitive–affective states. PPG measures blood volumne changes and features such as pulse transit time, pulse amplitude and pulse wave variability reflect changes in cardiovascular and autonomic nervous system activity \cite{Intro_4_PPG}.

Several publicly available datasets have been published that include physiological and emotional response data. Some of these datasets are WESAD \cite{wesad}, DREAMER \cite{DREAMER}, DEAP \cite{deap}, AMIGOS \cite{amigos}, K-EmoCon \cite{k_emocon} and CLAS \cite{clas}. These datasets typically include signals such as ECG, PPG, EDA, and EEG recorded under controlled emotional stimulus conditions. A wide range of machine learning algorithms such as Naive Bayes, SVM (Support vactor machine), ANN (Artificial neural network), RF (Random Forest) and Ensemble algorithms have been explored in the literature for emotion classification, targeting different emotional representations such as discrete emotion categories and continuous valence–arousal dimensions \cite{Related_work1, Related_work2, Related_work3, Related_work5, Related_work6, Related_Work7, Related_Work8} with classification accuracy for discrete emotions ranging between 39.2\% and 56.4\% and for valence/arousal between and 48\% and 81\%. 

This study presents a newly collected dataset comprising of physiological signals recorded while participants are watching emotion-eliciting stimuli. The study also aims to assess the impact of different targets on emotion classification performance using traditional machine learning algorithms. In line with this objective, the present study addresses the following research questions:
\begin{itemize}
    \item How feasible it is to use ECG and GSR data acquired from Shimmer devices for developing emotion recognition models?
    \item Which target representation -- valence, arousal or discrete emotions classes -- leads to better emotion classification performance?
    \item Can light-weight/traditional ML models achieve good accuracy for emotion recognition?
\end{itemize}

\section{Experimental Framework}\label{ExperimentalSetup}
This section describes the experimental framework that aims to collect the physiological signals associated with different emotional stimuli in controlled environment. The subsections below describe the participant details, the design and arrangements of the emotional stimuli, physiological signals measured and devices used and the procedure followed during data collection.
\subsection{Participants}
A total of 22 participants were recruited for this study comprising of 15 Males and 7 Females. Participants were recruited through an online questionnaire. Participants were categorized into three age groups: 20-35 (young adults), 36-50 (Adults) and 50+ (middle aged). In terms of nationality, participants included 11 international and 11 Finnish subjects. The details regarding the distribution as per age, gender and nationality are given in Table \ref{tab:demographics}. Inclusion criteria required participants to be above 18  years of age and with no diagnosed long-term illnesses that affect the functioning of the central nervous system. All participants were provided detailed documents regarding the study and written informed consent was obtained prior to the study. The study was approved by the Human Sciences Ethics Committee of the Universities in Satakunta, Finland (approval ID: 16.05.2025).
\begin{table}
\centering
\caption{Participant Demographics}
\label{tab:demographics}
\begin{tabular*}{\columnwidth}{|p{0.6in}|p{1.2in}|p{0.5in}|p{0.5in}|}
\hline
\textbf{Variable} & \textbf{Category} & \textbf{Count(in number)} & \textbf{Count(in \%)} \\
\hline
Gender & Male & 15 & 68.1\%\\
        \cline{2-4}
       & Female & 7 & 31.8\%\\
\hline
Age group & 20-35 (young adults) & 14 & 63.6\%\\
          \cline{2-4}
          & 36-50 (adults)& 5 & 22.7\%\\
          \cline{2-4}
          & 50+ (middle aged)& 3 & 13.6\%\\
          \cline{2-4}
\hline
Nationality & Finnish & 11 & 50\%\\
            \cline{2-4}
            & International & 11 & 50\%\\
\hline
\end{tabular*}
\end{table}

\subsection{Emotional Stimuli Design}
 Emotional stimuli were designed to evoke a range of emotions across different valence and arousal levels. The stimuli consisted of short video clips selected from \cite{video_clip_paper}. Out of the 18 clips, 9 clips were selected to represent positively valenced emotions such as amusement, excitement, and happiness and negative emotions such as anger, disgust, fear, sadness, and surprise. Neutral clips such as calmness were considered to have no valence. These movie clips were pre-annotated by over 300 participants and have been used by the DREAMER \cite{DREAMER} dataset for emotion recognition task. Duration of the movie clips representing emotional stimuli varied between 2 minutes to 6 minutes. These emotional stimuli were arranged such that after each stimulus, a neutral clip of about 60 sec was shown to the participants to reduce the emotional carryover effect from the previous stimulus. After each emotional stimulus, participants were displayed a self-assessment questionnaire containing evaluation of discrete emotions as well as valence and arousal according to the Self-Assessment Manikin (SAM) test. The ratings scale allows to collect the subjective response from participants after every clip to verify that the emotional stimuli elicited the intended emotions. The entire experimental protocol was implemented as a web-based system developed using Python that controlled stimulus presentation, video start and end timestamp marking and recording participants' subjective responses.
 
\subsection{Acquisition of Physiological Signals}
To capture the autonomic responses from the participants while they are watching movie clips, physiological responses such as ECG (Electrocardiogram) and GSR(Galvanic Skin Response) were recorded using wearable devices. ECG signals were recorded using the Shimmer 3 EXG module with three-lead configuration and  4 electrodes placed on the chest of the subject. GSR were recorded using the Shimmer 3 GSR module from the subject's fingers of  the non-dominant hand. Both these signals were sampled at 256 Hz. ECG and GSR were recorded to capture the heart rate, heart rate variability and sympathetic nervous system activity, respectively, during emotional arousal. All signals were synchronized according to the stimulus presentation time.

\subsection{Experimental Procedures}
The experimental session began with a briefing about the study, the video protocol, sensors used and placement of electrodes as well as the assessment form. Participants signed a written consent form before the beginning of the session. Participants were seated in a quite, temperature/light-controlled room with minimal external distractions. The devices were then calibrated and attached to the participant. The video protocol starts with a neutral baseline clip of about 60 seconds followed by a series of emotional stimuli in the following order: Amusement, Anger, Calm, Disgust, Excitement, Fear, Happy, Sad, Surprise. The order of showing the clips was kept the same throughout. After every emotional stimulus participants filled the self-assessment form followed by a neutral clip of 60 seconds. The entire video protocol took approximately 45 minutes. Upon completion of the session, sensors were removed and participants were debriefed.

\section{Data Analysis Framework}\label{Implementationdetails}
This section presents the data analysis framework used for analyzing the physiological signals for emotion recognition using the acquired data from 22 participants. The framework begins with data pre-processing and labeling, followed by signal pre-processing and feature extraction. Extracted features were used to train machine learning models for emotion classification. Model performance is evaluated using repeated stratified cross-validation. The study utilized multiple hardware sensors and software tools for data acquisition, preprocessing, feature extraction, and model development. Physiological signals were collected using devices mentioned in section \ref{ExperimentalSetup} , while data preprocessing and feature extraction were performed in python using scipy and neurokit2 libraries for ECG and GSR respectively. Machine learning analysis and visualization were carried out using Python libraries such as NumPy, Pandas, Scikit-learn, and Matplotlib. Additional synchronization and signal processing steps were implemented to standardize the collected data. The subsections below describe in detail the step-by-step process.

\subsection{Data Pre-processing and Exploratory Data Analysis}
This section describes the pre-processing steps applied to the raw physiological signals obtained from sensors and the exploratory data analysis conducted on physiological signals and subjective data collected from participants. The raw ECG and GSR data were obtained in comma separated value (CSV) file format. The raw ECG file contains the data from three leads, out of which we selected the LL–RA (Left Leg – Right Arm) ECG lead configuration because it captures cross-sectional cardiac electrical activity and provides robust R-wave detection for reliable heart rate and heart rate variability analysis. Similarly, from raw GSR file we used the data from skin conductance variable. The start and end times  of the movie clips (emotional stimuli) were also recorded in CSV file format through the video protocol. Using this information, the raw ECG and GSR data were annotated as per the timestamp information for video start and end times. We also marked the timestamps information for neutral clips for visualization purposes. The next step was signal filtering and noise removal. We used band-pass filter with the passband from 0.5 Hz to 45 Hz for ECG. During annotations, it was found that due to sensor calibration issues, sampling frequency was different for some of the participants, so additional interpolation step was carried out in these cases.  

We conducted exploratory data analysis on objective data (data from ECG and GSR) and subjective data (data from self-assessment forms displayed after each emotional stimuli) to evaluate visually signal integrity and response reliability. For the purpose of visualizing ECG and GSR data we extracted heart rate and skin conductance level corresponding to the different video segments for all participants.
\begin{figure*}[t]
  \centering
  \includegraphics[width=\textwidth]{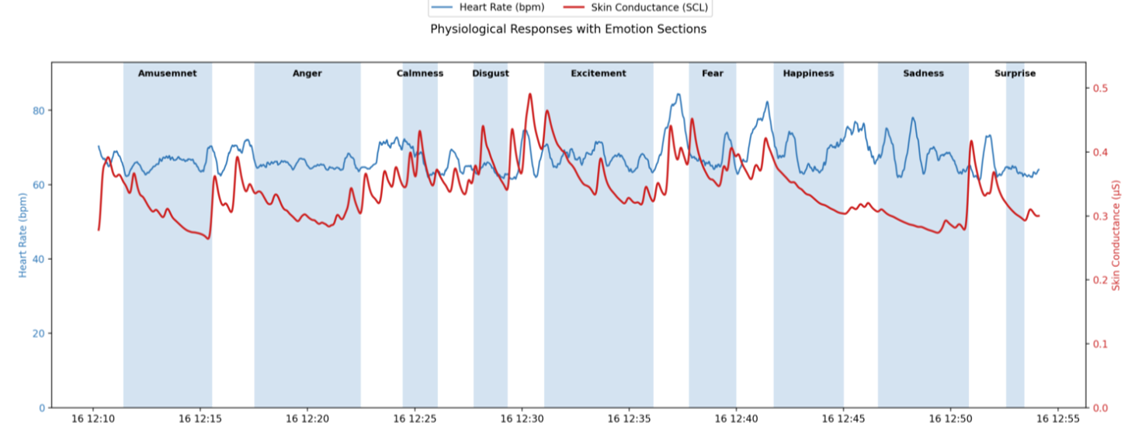}  
  \caption{Heart rate and skin conductance level over video segments}
  \label{fig:HR_SCL}
\end{figure*}
Figure \ref{fig:HR_SCL} shows the heart rate and skin conductance level for a selected participant throughout the whole recording. The video segments are marked with blue shaded region. It can be seen that heart rate (blue line) is pretty much constant but there are some minor peaks after excitement, fear and during the sadness clip. For skin conductance, major peaks were seen after disgust and excitement video clips. It was also seen that skin conductance became normal after the fear clip and during the happy clip but it was further lower during the sad video clip. Similar trends in heart rate and skin conductance level were observed for most of the participants. Additionally, it was also seen that for some participants skin conductance level was higher during the questionnaires except for the intense stimuli segments. Furthermore, the carry-over effect from intense emotional stimuli was also observed in the skin conductance for a few participants.

\begin{figure}
  \includegraphics[width=0.9\columnwidth]{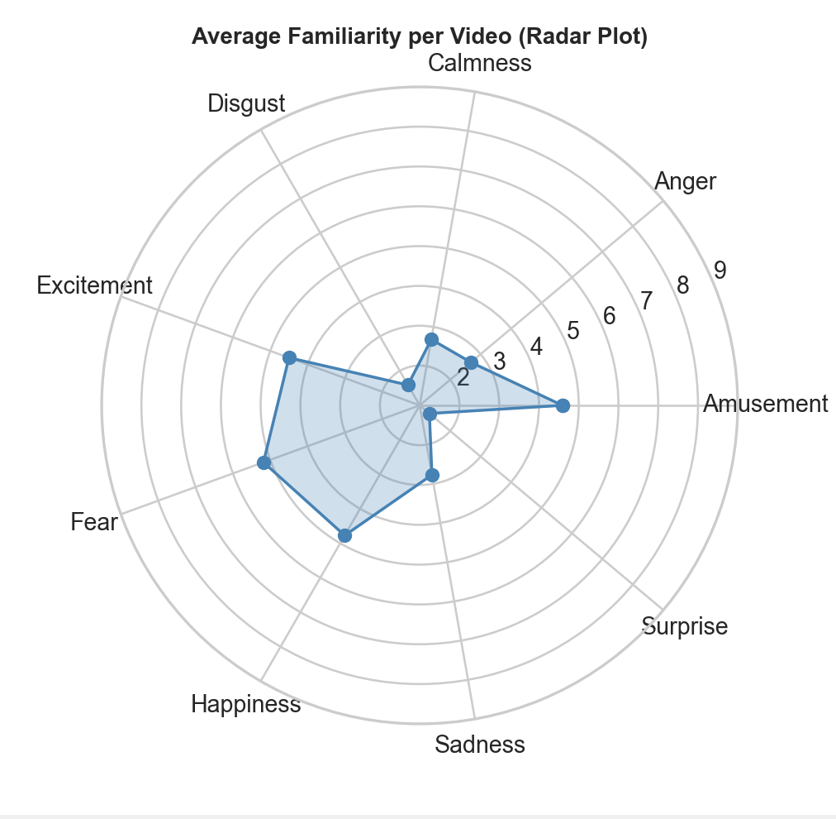}
  \caption{Participant's familiarity with the video clips}
  \label{fig:familiarity}
\end{figure}
\begin{figure}
  \includegraphics[width=0.9\columnwidth]{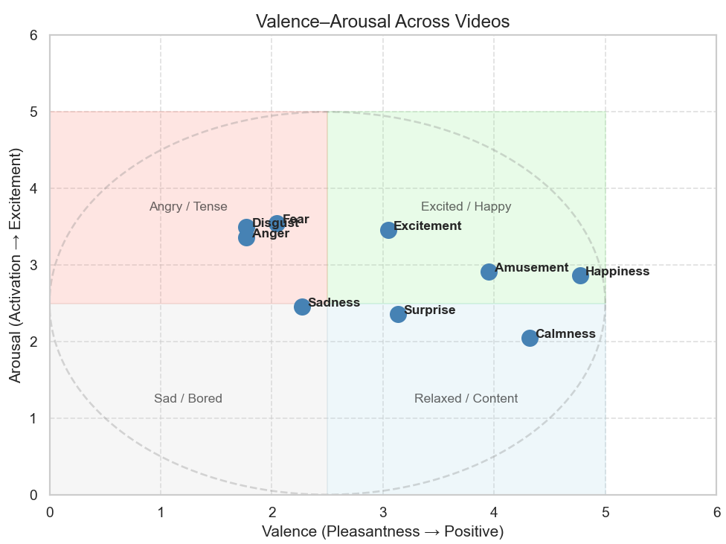}
  \caption{Participant's Valence and Arousal ratings}
  \label{fig:valence_arousal}
\end{figure}

We also conducted subjective analysis using valence and arousal ratings obtained from the Self-Assessment Manikin (SAM) scale, along with familiarity ratings reported by participants. Figure \ref{fig:familiarity} shows the radar chart indicating the all 22 participant's familiarity data with the video clips. Lower number shows lesser familiarity with the clip. On the average, the participants were less familiar with the disgust, calm, surprise and anger clips as compared to other clips and more familiar with the fear and happiness clips. Figure \ref{fig:valence_arousal} shows the valence--arousal ratings by participants (all 22 participants included) and to which region, as per Russells circumplex model of emotions \cite{Russell_circumplex_model}, the video clip belongs. As per these curves, except for sadness and surprise, all the videos align with the actual valence arousal space which shows that the stimuli elicited the targeted emotions.

\subsection{Feature Extraction} \label{sec:featureextraction}
This section describes the feature extraction process used to derive representative features from the pre-processed physiological signals for subsequent machine learning analysis. The modalities considered were ECG and GSR. We also used the self-reported valence and arousal values for the classification. Features were extracted from signal segments corresponding to each emotional stimulus. For extracting the features, the continuous ECG and GSR recordings were segmented into non-overlapping windows of $10$ seconds. We extracted time-domain and frequency-domain features from ECG and time-domain and phasic response features from GSR.

ECG signals reflect autonomic nervous system activity, therefore time-domain features such as heart rate, heart rate variability, and RR intervals were extracted to quantify physiological changes during emotional stimulation. Frequency-domain features, including LF (low-frequency) and HF (High-frequency) band power and the LF/HF ratio, were also computed to complement the time-domain measures. Table \ref{tab:featuresetECG} describes in detail the features extracted. Altogether, we obtained 12 features from the ECG. GSR signal represents skin's electrical conductivity and is primarily related to sympathetic (arousal) nervous system activity, making it a sensitive indicator of high-arousal emotions such as fear, anger, disgust, sadness or excitement. GSR signals have two components, Skin Conductance Level (SCL or tonic component) and Skin Conductance Response (SCR or phasic component). SCL tells about general arousal whereas SCR shows rapid peaks and instantaneous activity corresponding to emotional stimulus. To capture the insights from both components, Tonic and phasic component features described in Table \ref{tab:featuresetGSR} were computed from both components totaling to overall 13 features from GSR.
\begin{table}
\caption{Feature set extracted from ECG signals} 
\label{tab:featuresetECG}
\footnotesize\centering
\begin{tabular*}{\columnwidth}{|p{0.6in}|p{0.5in}|p{1.89in}|}
  \hline
  \textbf{Category} & \textbf{Feature} & \textbf{Description}\\
  \hline
  Time-domain &  Mean RR & Mean of the time interval between successive normal heart beats  (i.e. the RR-intervals) over a window\\
  \hline
  Time-domain &  STD RR & Standard deviation of the time interval between successive normal heart beats (i.e. the RR-intervals)\\
  \hline
  Time-domain &  Mean HR & Average heart rate over a window\\
  \hline
  Time-domain &  STD HR & Standard deviation of heart rate\\
  \hline
  Time-domain &  Minimum HR & Minimum value of heart rate over a window\\
  \hline
  Time-domain &  Maximum HR & Maximum value of heart rate over a window\\
  \hline
  Time-domain & RMSSD & The square root of the mean of the sum of the squares of differences between adjacent RR-intervals\\
  \hline
  Time-domain & PNN50 & The percentage of interval differences of successive RR-intervals greater than 50 ms\\
  \hline
  Frequency-domain & VLF (Very low-frequency) & Indicates slow, long-term regulatory activity of the autonomic nervous system, over a frequency range of 0.0033–0.04 Hz\\
  \hline
  Frequency-domain & LF & Indicates combined sympathetic and parasympathetic activity, over a frequency range of 0.04–0.15 Hz\\
  \hline
  Frequency-domain & HF & Indicates parasympathetic activity, over a frequency range of 0.15–0.4 Hz\\
  \hline
  Frequency-domain & LF/HF ratio & Ratio of LF and HF indicating balance between sympathetic and parasympathetic activity\\
  \hline
\end{tabular*}
\end{table}

\begin{table}
\caption{Feature set extracted from GSR signals}
\label{tab:featuresetGSR}
\footnotesize\centering
\begin{tabular*}{\columnwidth}{|p{0.6in}|p{0.5in}|p{1.89in}|}
  \hline
  \textbf{Category} & \textbf{Feature} & \textbf{Description}\\
  \hline
  Tonic-level &  Mean SCL & Average conductance over a window, reflecting general arousal\\
  \hline
  Tonic-level &  STD SCL & Standard deviation of conductance over a window, measures fluctuations\\
  \hline
  Tonic-level &  Minimum SCL & Measures lowest conductance level over a window\\
  \hline
  Tonic-level &  Maximum SCL & Measures lowest conductance level over a window\\
  \hline
  Tonic-level &  Skewness & Measures the asymmetry of the distribution\\
  \hline
  Tonic-level &  Kurtosis & Measures the tailedness or peakedness of the distribution\\
  \hline
  Tonic-level & Mean peak amplitude & Measures mean peak to peak amplitude over a window\\
  \hline
  Tonic-level & Mean slope & Measure rate of change over time, indicates gradual increase/decrease.\\
  \hline
  Tonic-level & Peaks & Measures sum of areas under the signal curve at each detected peak\\
  \hline
  Phasic-level & Number of SCR peaks & Count of distinct responses in a time window\\
  \hline
  Phasic-level & Mean SCR amplitude & Average skin conductance response amplitude in a window\\
  \hline
  Phasic-level & Mean rise time & Average time from onset to peak of response over a window\\
  \hline
  Phasic-level & Mean recovery time & Average time it takes for signal to fall back to 50\% after peak\\
  \hline
\end{tabular*}
\end{table}

\subsection{Emotion Recognition Model and Evaluation}
To classify emotional states using physiological signals, machine learning framework was implemented. In addition to the features described in section \ref{sec:featureextraction}, the subjective valence and arousal values were also used for classification. Valence and arousal are the most important dimensions for emotion recognition study. These dimensions enable the models to capture subtle variations especially with the mixed emotional states hence we thought it is worth exploring these features for machine learning modeling. We used several supervised learning models, including Support Vector Machines (SVM), Random Forest (RF), Extreme gradient boosting (XGBoost) and Neural network (NN). Since the data were obtained from limited set of participants, evaluating traditional ML models was the first choice rather than using complex deep learning models that could overfit. We used repeated stratified cross-validation to ensure that every fold fairly represents all emotion classes, since there were few emotion classes that were present in minority. This process was repeated 50 times for SVM, RF, XGBoost and 10 times for NN to provide a robust estimate of how well the model will perform on unseen data. Model performance was assessed using accuracy as the evaluation metrics. We also calculated feature importance metrics for top performing model to determine which physiological features contributed most to emotional state classification.

\section{Results}
\begin{figure}
  \includegraphics[width=0.9\columnwidth]{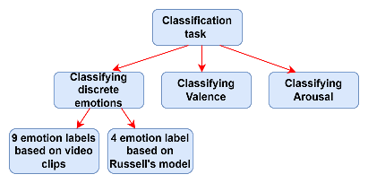}
  \caption{Classification tasks performed}
  \label{fig:classificationtask}
\end{figure}

This section presents the results of the analysis of physiological signals and subjective valence and arousal values for the emotion recognition task. It includes the performance of various machine learning algorithms on ECG feature set, GSR feature set and combined ECG and GSR feature set. We used several different classification setups as summarized in Figure \ref{fig:classificationtask}. 
We tried classifying discrete emotions with 9 emotion classes (similar to the emotions of the movie clips) and 4 emotion classes (combining the emotions that belong to the same quadrant of the Russell's circumplex model of emotions \cite{Russell_circumplex_model}). Since for emotion recognition studies, valence and arousal are considered as a crucial dimension and are associated closely with physiological responses, we have also tried classifying these dimensions. In here, we have used two cases of data, one with subjective data, which means classification includes the subjective measurements (such as valence and arousal in case of classifying emotions or discrete emotions in case of classifying valence and arousal) and one without subjective data which means only physiological data is used for classification.

Tables \ref{tab:acc_ECG_classification_cases},\ref{tab:acc_GSR_classification_cases} and \ref{tab:acc_ECG_GSR_classification_cases} present the results of machine learning models using only ECG sensor data, GSR sensor data and ECG+GSR sensor data respectively for all the four type of classification tasks and for the two sets of data. In tables, estimating discrete emotions with subjective data means that in addition to physiological signal features, valence and arousal were also used as inputs and without subjective data means only physiological signals were used as inputs. Similarly, estimating valence and arousal with subjective data means discrete emotion classes were used as inputs in addition to physiological signals and without subjective data means only physiological signal features were used. For only ECG data, it can be seen that when subjective data is considered, highest accuracy of 84.7\% is obtained by the random forest algorithm for 4 emotion classes. However, when only physiological data is used, Neural network performed better with 47.6\% accuracy for 4 emotion classes. For only GSR data, random forest obtained 86.15\% accuracy when subjective data was included and XGBoost obtained 51.1\% accuracy without subjective data. In this case, the classification task involving predicting discrete emotions with 4 classes showed highest accuracy. When both ECG and GSR data were used, again random forest algorithm with 89.8\% accuracy (subjective data included) performed better for classifying discrete emotions. However, when only physiological data was used, random forest showed good performance for classifying arousal with 53.4\% accuracy. Additionally, XGBoost (52.4\%) and random forest (53.0\%) also showed good performance for classifying discrete emotions with 4 classes.
\begin{table*}
\caption{Classification accuracy (\%) for ECG feature set across different scenarios} 
\label{tab:acc_ECG_classification_cases}
\footnotesize\centering
\begin{tabularx}{\textwidth}{|
    p{0.55in}|    
    *{4}{p{0.63in}|}  
    *{4}{p{0.63in}|}  
}
\hline
\textbf{Algorithm} &
\multicolumn{4}{c|}{\textbf{With subjective data}} &
\multicolumn{4}{c|}{\textbf{Without subjective data}} \\
\hline

{} &
{\centering \textbf{Discrete emotions (9 classes)}} &
{\centering \textbf{Discrete emotions (4 classes)}} &
{\centering \textbf{Valence classification}} &
{\centering \textbf{Arousal classification}} &
{\centering \textbf{Discrete emotions (9 classes)}} &
{\centering \textbf{Discrete emotions (4 classes)}} &
{\centering \textbf{Valence classification}} &
{\centering \textbf{Arousal classification}} \\
\hline
RF &
66.0\% & \textbf{84.7}\% & 78.3\% & 72.2\% & 21.0\% & 45.1\% &33.8\% & 39.1\%\\
\hline
XGBoost &
58.0\% & 83.5\% & 75.3\% & 65.8\% & 21.3\% & 46.9\% &35.7\% & 39.8\%\\
\hline
SVM &
57.3\% & 79.1\% & 67.3\% & 64.9\% & 15.6\% & 32.8\% &28.9\% & 33.5\%\\
\hline
ANN &
46.0\% & 76.8\% & 40.3\% &41.9\% & 18.3\% & \textbf{47.6}\% & 31.3\%& 36.1\%\\
\hline
\end{tabularx}
\end{table*}

\begin{table*}
\caption{Classification accuracy (\%) for GSR feature set across different scenarios} 
\label{tab:acc_GSR_classification_cases}
\footnotesize\centering
\begin{tabularx}{\textwidth}{|
    p{0.55in}|    
    *{4}{p{0.63in}|}  
    *{4}{p{0.63in}|}  
}
\hline
\textbf{Algorithm} &
\multicolumn{4}{c|}{\textbf{With subjective data}} &
\multicolumn{4}{c|}{\textbf{Without subjective data}} \\
\hline

{} &
{\centering \textbf{Discrete emotions (9 classes)}} &
{\centering \textbf{Discrete emotions (4 classes)}} &
{\centering \textbf{Valence classification}} &
{\centering \textbf{Arousal classification}} &
{\centering \textbf{Discrete emotions (9 classes)}} &
{\centering \textbf{Discrete emotions (4 classes)}} &
{\centering \textbf{Valence classification}} &
{\centering \textbf{Arousal classification}} \\
\hline
RF &
72.4\% & \textbf{86.1}\% & 83.8\% & 80.7\% & 25.6\% & 44.0\% &34.2\% & 39.0\%\\
\hline
XGBoost &
66.9\% & 84.6\% & 76.4\% & 70.4\% & 32.4\% & \textbf{51.1}\% &38.8\% & 45.0\%\\
\hline
SVM &
60.0\% & 77.9\% & 73.0\% & 67.0\% & 17.6\% & 40.9\% &27.6\% & 29.6\%\\
\hline
ANN &
49.0\% & 77.9\% & 44.4\% & 47.4\% &20.2\% &47.6\% & 28.1\% & 36.3\%\\
\hline
\end{tabularx}
\end{table*}

\begin{table*}
\caption{Classification accuracy (\%) for ECG+GSR feature set across different scenarios} 
\label{tab:acc_ECG_GSR_classification_cases}
\footnotesize\centering
\begin{tabularx}{\textwidth}{|
    p{0.55in}|    
    *{4}{p{0.63in}|}  
    *{4}{p{0.63in}|}  
}
\hline
\textbf{Algorithm} &
\multicolumn{4}{c|}{\textbf{With subjective data}} &
\multicolumn{4}{c|}{\textbf{Without subjective data}} \\
\hline

{} &
{\centering \textbf{Discrete emotions (9 classes)}} &
{\centering \textbf{Discrete emotions (4 classes)}} &
{\centering \textbf{Valence classification}} &
{\centering \textbf{Arousal classification}} &
{\centering \textbf{Discrete emotions (9 classes)}} &
{\centering \textbf{Discrete emotions (4 classes)}} &
{\centering \textbf{Valence classification}} &
{\centering \textbf{Arousal classification}} \\
\hline
RF &
75.6\% & \textbf{88.1\%} & 85.3\% & 86.4\% & 32.1\% & \textbf{53.0}\% &45.5\% & \textbf{53.4\%}\\
\hline
XGBoost &
69.3\% & 87.8\% & 79.8\% & 77.8\% & 33.6\% & 52.4\% & 44.1\% & 52.1\%\\
\hline
SVM &
54.1\% & 75.7\% & 58.4\% & 63.6\% & 22.3\% & 43.4\% &33.7\% & 41.2\%\\
\hline
ANN &
51.6\% & 81.3\% & 46.5\% & 49.3\% & 21.6\% &48.1\% & 33.1\% & 39.8\%\\
\hline
\end{tabularx}
\end{table*}

\begin{figure}
  \centering
  \includegraphics[width=0.9\columnwidth]{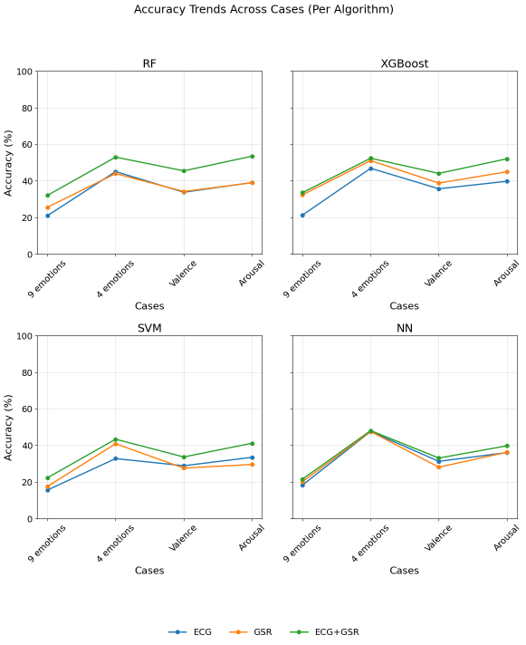}  
  \caption{Accuracy trend per algorithms}
  \label{fig:Accuracy_per_algorithm_trend}
\end{figure}

\begin{figure}
  \centering
  \includegraphics[width=1.1\columnwidth]{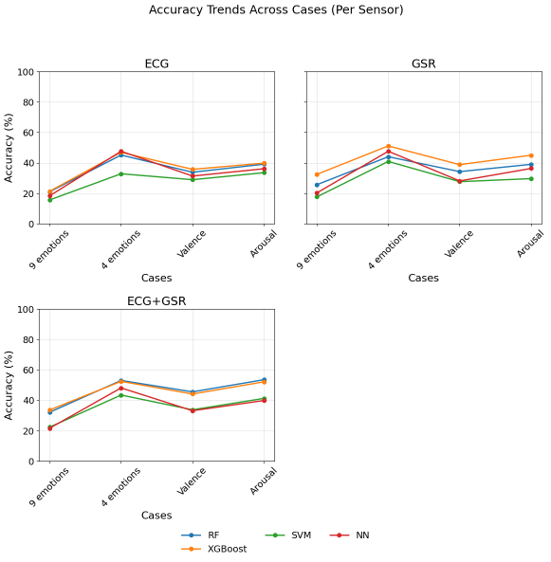}  
  \caption{Accuracy trend per sensor}
  \label{fig:Accuracy_per_sensor_trend}
\end{figure}

Figures \ref{fig:Accuracy_per_sensor_trend} and \ref{fig:Accuracy_per_algorithm_trend} show how accuracy varies across sensors and across algorithms for different classification cases respectively. Although classification accuracies using subjective features was also evaluated to examine the influence of subjective measures but for performance comparison, we have compared accuracies with only physiological data. It can be seen from the figures that for all algorithms and all classification cases, the combination of ECG+GSR performed better compared to the individual ECG and GSR sensors. For all the algorithms, the performance of valence classification and discrete emotions for 4 classes have shown better accuracies than valence classification and discrete emotion classification with 9 classes. The classification performance of discrete emotions with 4 classes has shown highest accuracy, maximum being 53.0\% for ECG+GSR combination using random forest. This could be attributed to the fact that the inter-class overlaps are reduced with 4 emotions and physiological signals are more effective in capturing broader affective states than fine-grained emotional distinctions. This performance accuracy obtained is substantially better than the random accuracy, which is 35\% (as per the class distribution of 4 emotions), indicating that the model learned meaningful patterns from the physiological signals without any subjective reference. Additionally, a significant increase (~10\%) in the performance of the arousal classification was observed for ECG+GSR as compared to the individual sensors ECG and GSR. The classification accuracy of 53.4\% was obtained for arousal classification using random forest algorithm. This result could be because the arousal reflects the intensity of emotions, which is strongly expressed by physiological data than valence. Moreover, the fusion of complementary signals ECG and GSR improves the class separability resulting in higher arousal classification. Figure \ref{fig:Accuracy_per_algorithm_trend} compares the performance of the different classifiers. If we compare different feature sets then Random forest, XGBoost and Neural network performed better for ECG feature set than SVM algorithm whereas in case of GSR feature set random forest have shown good performance over other algorithms. For combined feature set, Random forest and XGBoost showed good performance. This shows that random forest consistently performed better for all the feature sets.

\section{Conclusion}
This study investigated the performance of various machine learning algorithms for emotion recognition based on discrete emotions, valence and arousal as target variable. We also explored emotion class granularity by evaluating 9 class (based on video clips) and 4 class (based on Russell's circumplex model) classification. Emotional responses were elicited through designed video protocol consisting of video clips (from DREAMER dataset) and neutral clips. Physiological responses such as ECG, and GSR were collected along with subjective data to ensure reliable emotion labeling. For the scope of this study, ECG and GSR signals were analyzed through systematic pre-processing, exploratory analysis, and feature extraction and machine learning analysis. Machine learning models were evaluated using repeated stratified cross-validation. The results showed that the best performance was obtained for classifying arousal (53.4\%) and discrete emotions with 4 classes (53.0\%) for ECG+GSR feature set with Random Forest algorithm. For individual sensors, XGBoost for GSR (51.1\%) and neural network for ECG (47.6\%) showed good performance for classifying discrete emotions with 4 classes. From the analysis it can be inferred that classification accuracies obtained using traditional models were better than the random accuracies which shows that model has learned meaningful patterns from the data especially only physiological data. These findings suggest that it is feasible to use physiological responses obtained Shimmer devices such as ECG and GSR for recognizing emotions with discrete emotions and valence/arousal. Future work could focus on incorporating the data from more participants along with enhancing the feature set by including the features from EEG and PPG sensors and replicating the model in real world environments.

\bibliographystyle{IEEEtran}
\bibliography{Bibliography}{}
\end{document}